# Topological phase transition and interface states in hybrid plasmonic-photonic systems

Lixin Ge[1], Liang Liu[1], Meng Xiao[2], Guiqiang Du[3], Lei Shi[4], Dezhuan Han[1],
C. T. Chan[5] and Jian Zi[4]

[1] Department of Applied Physics, Chongqing University, Chongqing 401331, China
[2] Department of Electrical Engineering, and Ginzton Laboratory, Stanford University, Stanford, California 94305, USA
[3] School of Space Science and Physics, Shandong University at Weihai, Weihai 264209, China
[4] Department of Physics, Key laboratory of Micro and Nano Photonic Structures (MOE), and Key Laboratory of Surface Physics, Fudan University, Shanghai 200433, China
[5] Department of Physics and Institute for Advanced Study, Hong Kong University of Science and Technology, Hong Kong, China

Email: dzhan@cqu.edu.cn, jzi@fudan.edu.cn

**Abstract**
The geometric phase and topological property for one-dimensional hybrid plasmonic-photonic crystals consisting of a simple lattice of graphene sheets are investigated systematically. For transverse magnetic waves, both plasmonic and photonic modes exist in the momentum space. The accidental degeneracy point of these two kinds of modes is identified to be a diabolic point accompanied with a topological phase transition. For a closed loop around this degeneracy point, the Berry phase is π as a consequence of the discontinuous jump of the geometric Zak phase. The wave impedance is calculated analytically for the semi-infinite system, and the corresponding topological interface states either start from or terminate at the degeneracy point. This type of localized interface states may find potential applications in photonics and plasmonics.

______________________________________________________________________________________________________

## 1. Introduction

Recently, the concept of "topology" has attracted much research interest in photonic systems [1]. For a two-dimensional periodic system, the Brillouin zone forms a torus. The integral of the Berry curvature over this closed torus is topologically invariant and quantized as an integer, termed Chern number [2]. In photonic systems, by either breaking time-reversal symmetry or introducing spin-orbital couplings, non-zero Chern numbers can appear and a robust one-way transportation can be realized [1,3-4]. For one-dimensional photonic systems, the topological phase, characterized by the "Zak phase" [5], has also been investigated recently [6]. The relationship between Zak phases and surface impedance, the "bulk-interface correspondence", was derived analytically in one-dimensional photonic crystals (PCs). The measurements of the Zak phases for one-dimensional PCs have also been carried out recently [7].

For PCs composed of both plasmonic and dielectric materials, not only plasmonic modes but also photonic modes exist [8-9]. Plasmonic modes have the advantage of field confinements in the subwavelength scale, but unfortunately with an unavoidable intrinsic loss. In contrast, for the photonic modes the loss is extremely small but the scale of the field confinement is comparable to the wavelength. Strikingly, the hybridization of these two kinds of modes can provide a scheme to combine both advantages. Graphene, being only one atomic thick, can support plasmonic modes

[10]. Graphene plasmons possess many remarkable features including deep subwavelength, low loss, and high tunability, serving as a promising platform for strong light-matter interaction in terahertz and infrared frequencies [11-12].

In this paper, we study systematically the geometric phase and topological property for a hybrid plasmonic-photonic system. Firstly, we take a simple 1D lattice of graphene sheets as an example. For transverse magnetic modes, there exists an accidental degeneracy point stemming from the hybridization of plasmonic and photonic modes. As the parallel wavevector component is equal to a critical value, the photonic band gap is closed at a Diabolic point [13]. At this point the band inversion of plasmonic and photonic modes takes place and a topological phase transition occurs. Therefore, the quantized Zak phases of the first and second photonic bands undergo a discontinuous jump. Around this transition point, the Berry phase is π, implying that this accidental degeneracy is non-trivial [14]. Topological interface states are demonstrated for a semi-infinite plasmonic-photonic crystal. The dispersion of these interface states depends on the nature of the ambient medium (dielectric or plasmonic), which can also be interpreted by the wave impedance analytically, manifesting the "bulk-interface correspondence" in this system.

## 2. Band structures and geometric phases of a simple lattice of graphene sheets

The system in this study is schematically shown in figure 1(a). The graphene sheets are embedded periodically in a host medium with a dielectric constant $\varepsilon_h$. The optical conductivity of graphene can be given by the following relation within the random phase approximation [15-16]:

$$\tilde{\sigma} \equiv \frac{\sigma}{\varepsilon_0 c} = 4\alpha \frac{i}{\Omega} + \pi\alpha \left[ \theta(\Omega - 2) + \frac{i}{\pi} \ln \left| \frac{\Omega - 2}{\Omega + 2} \right| \right], \quad (1)$$

where $\varepsilon_0$ is the vacuum permittivity, $c$ is the speed of light in vacuum, $\alpha$ is the fine structure constant, $\theta(x)$ is the Heaviside step function, and $\Omega \equiv \hbar\omega/E_F$ is the dimensionless frequency (here, $\omega$ is the angular frequency, $\hbar$ is the reduced Planck constant, and $E_F$ is the Fermi energy).

There exist two independent polarizations: transverse magnetic (TM) and transverse electric (TE) ones with the magnetic and electric field polarized along the y-axis, respectively. The dispersion relation for these two modes can be solved by a transfer matrix method [17-18] given by:

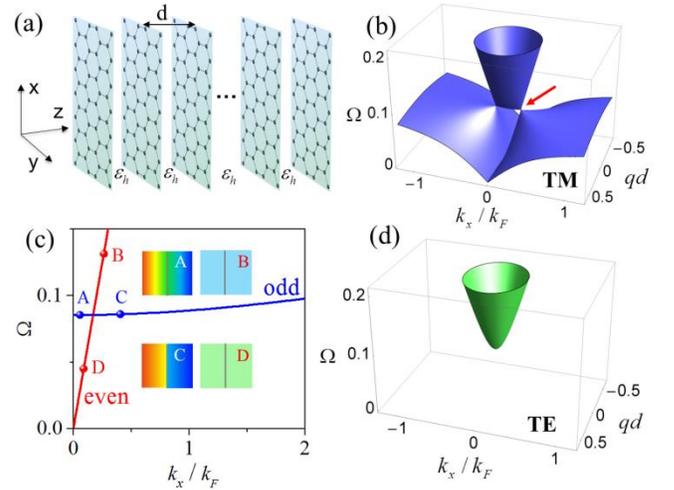

**Figure 1.** (a) Schematic view of a simple lattice of graphene sheets embedded in a host medium with dielectric constant $\varepsilon_h$. (b) and (d) Photonic band structures for the TM and TE modes, respectively. In (b), the accidental degeneracy point is indicated by the red arrow. (c) Dispersion for TM modes at $q=0$. In the inset, the distributions of the magnetic field in a unit cell are shown for points A, B, C, and D with the gray lines representing the graphene sheet. The parameters used are $\varepsilon_h=4$, $d=1.0\,\hbar c/E_F$, and $k_F=E_F/(\hbar c)$.

_________________________________________________

$$\text{TM:} \quad \cos(qd) = \cos(k_z d) - \frac{i\tilde{\sigma}k_z}{2\varepsilon_h k_0}\sin(k_z d), \quad (2.1)$$

$$\text{TE:} \quad \cos(qd) = \cos(k_z d) - \frac{i\tilde{\sigma}k_0}{2k_z}\sin(k_z d), \quad (2.2)$$

where $q$ is the Bloch wavevector, $d$ is the period, $k_z = \sqrt{\varepsilon_h k_0^2 - k_x^2}$ is the normal wavevector, $k_x$ is the parallel component, and $k_0=\omega/c$ is the wavevector in vacuum. The corresponding photonic band structures for TM and TE modes are shown in figures 1(b) and 1(d), respectively. The photonic band structures for the TM and TE mode are quite different. For the TM mode, the first photonic band is plasmonic from graphene plasmons and the higher bands are photonic. Interestingly, there exists a band-crossing point, arising from the accidental degeneracy of plasmonic and photonic modes [19]. This accidental degeneracy point, denoted by $(k_c, \Omega_c)$, is the diabolic point of the band structure [13]. While for the TE mode, there is no such band-crossing point due to the absence of plasmonic modes.

In order to study the geometric phase of the photonic band structures, the Bloch wave-function $H_{q,k_x}(z) = u_{q,k_x}(z)e^{iqz}$ is obtained by using the transfer matrix method [6,17-18], where $u_{q,k_x}(z)$ is a periodic function. In the unit cell, we choose the origin at $z=0$ as an inversion center, at which the graphene sheet

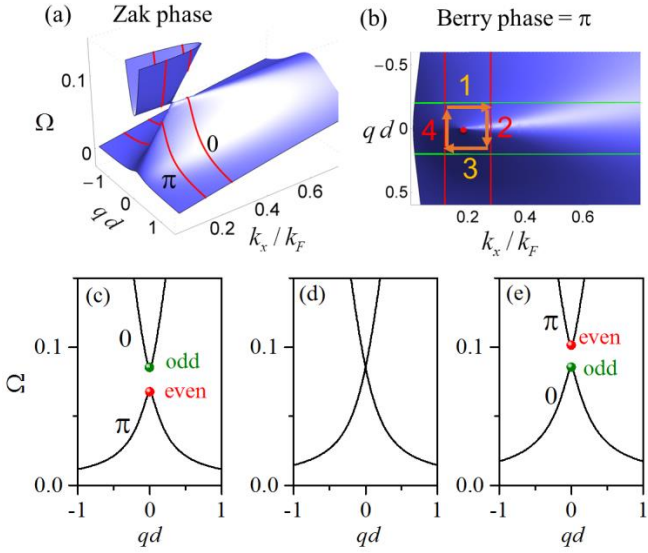

**Figure 2.** (a) Zak phases. The corresponding integral paths are shown by red lines. For the plasmonic mode (lower band), the Zak phase has a discontinuous jump from π to 0 as $k_x$ crosses the critical value $k_c$. (b) Berry phase. A closed integral path around the accidental degeneracy point is shown by orange arrows. (c), (d), and (e) show the band structures with $k_x$ being $0.8k_c$, $k_c$ and $1.2k_c$, respectively. The symmetry properties of the wave-function at the Γ point are also highlighted. Band inversion occurs at $k_x=k_c$. The parameters $\varepsilon_h$ and $d$ are the same as those in figure 1.

___

locates. It is found that $H_{q,k_x}(z)$ is an odd function of $z$ for the band edge states with $q=\pi/d$. We calculate $H_{q,k_x}(z)$ from the transfer matrix method for $\pi/d \geq q \geq 0$, while we choose that $H_{-q,k_x}(z) = -H_{q,k_x}(-z)$ for $-\pi/d \leq q < 0$. It is easy to verify that this Bloch wave-function satisfies the periodic gauge $H_{q,k_x}(z) = H_{q+2\pi/d,k_x}(z)$ automatically [6]. In figure 1(c), we plot the band structure for $q=0$, which is two crossing lines consisting of a band with $k_x = \sqrt{\varepsilon_h}k_0$ (red line) and a less dispersive band (blue line). The red band is just the light line, while the blue band basically arises from the plasmonic modes. On the blue line, $H_{q=0,k_x}(z)$ is an odd function of $z$ as exemplified by the state C, in which a discontinuous jump of the magnetic field can be observed, manifesting the characteristics of plasmons. While for the light line, $H_{q=0,k_x}(z)$ is an even function illustrated by the state B and D.

If the parallel component $k_x$ is fixed, the Zak phase can be defined by [5-6]:

$$\theta^{Zak}_{k_x} = i \int\int_{\text{BZ unit cell}} \varepsilon(z) u^*_{q,k_x} \partial_q u_{q,k_x} dz dq, \qquad (3)$$

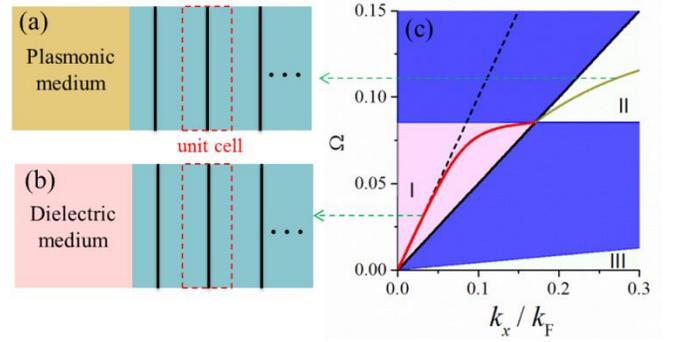

**Figure 3.** Dispersion of topological interface states for different ambient media. The semi-infinite structures are sketched in (a) and (b) for the plasmonic and dielectric medium, respectively. (c) The dispersion of topological interface states, calculated by Im($Z_s$)+Im($Z_a$)=0 analytically. The dark yellow and red lines correspond respectively to the plasmonic and dielectric ambient medium, as indicated by the arrows. The projected band structure for the periodic array of graphene sheets is also shown by the blue region. Region I, II and III are the forbidden band gaps with purely imaginary wave impedance $Z_s$. The dashed line stands for the light cone of the dielectric medium $\varepsilon_a$. Here, the parameters $\varepsilon_h$ and $d$ are the same as those in figure 1.

___

where $\varepsilon(z)=\varepsilon_h$ is the relative permittivity of the host medium. For convenience, the integration over the unit cell is written in a simple form $\langle u_{q,k_x} | \partial_q | u_{q,k_x} \rangle$ with the bra-ket notation. It can be proved directly that $\langle u_{q,k_x} | \partial_q | u_{q,k_x} \rangle$ is an odd function of $q$. Two examples of the integral path of the Zak phase are shown by the red lines in figure 2(a). By using Pancharatnam's discrete approach [20], the Zak phase is given by:

$$\theta^{Zak}_{k_x} = \left( \int_{-\pi/d}^{0^-} + \int_{0^+}^{\pi/d} dq \right) \langle u_{q,k_x} | \partial_q | u_{q,k_x} \rangle - \text{Im} \ln \langle u_{q=0^-,k_x} | u_{q=0^+,k_x} \rangle. \quad(4)$$

The first term vanishes since the integrand is odd in $q$, and the only contribution for the Zak phase comes from the second term in Eq. (4) [6]. Therefore, for the lower band with $k_x > k_c$, the Zak phase is 0 since $\langle u_{q,k_x} | \partial_q | u_{q,k_x} \rangle$ is continuous at $q=0$. Whereas for $k_x < k_c$, there is a discontinuous jump at $q=0$, and the corresponding Zak phase is π. The symmetry properties of the Bloch wave-function at $q=0$ and the corresponding Zak phase are shown in figures for different $k_x$. Analogous to that in one-dimensional dielectric PCs [6], Zak phases can also be determined by the symmetry properties of the band edge states. As mentioned above, the wave-functions are always odd functions of $z$ for the band edge states at $q=\pm\pi/d$. For $k_x=0.8\ k_c$ at the lower band, the Zak phase is π since the symmetry of the Bloch state at the Γ point is even. The photonic band gap closes at $k_x=k_c$ which gives rise to a diabolical point. For $k_x>k_c$, the gap reopens and the symmetries properties of band edge states are inverted. A topological phase transition takes place at $k_x=k_c$.

In two-dimensional photonic systems, the Berry phase can be used to identify the Dirac or Dirac-like spectrum [14]. In our system, Berry phase is also well defined for one-dimensional PCs if we incorporate the parallel wavevector component $k_x$:

$$\gamma = i \oint_C \varepsilon \left( \langle u_{q,k_x} | \partial_q | u_{q,k_x} \rangle, \langle u_{q,k_x} | \partial_{k_x} | u_{q,k_x} \rangle \right) \cdot (dq, dk_x)^T . \quad (5)$$

We choose an integral path around the degeneracy point as shown by the arrows in figure 2(b). For simplicity, we choose $q$ to be a constant for paths 1 and 3, while $k_x$ to be a constant for paths 2 and 4. It is easy to show that $\langle u_{q,k_x} | \partial_{k_x} | u_{q,k_x} \rangle$ is an even function of $q$. If we do a one-to-one mapping of $q \to -q$ with $k_x$ kept the same for path 1 and 3, the integral over these two paths cancels each other. For paths 2 and 4, the integral in Eq. (5) can be calculated similar to that in Eq. (4) for the Zak phase since $k_x$ is fixed along each path. The integral over path 2 is zero, while it is $\pi$ for path 4 as there exists a discontinuous point for the lower band at $q=0$. Thus, the Berry phase for this closed path around the degeneracy point is precisely $\pi$.

## 3. Topological interface states

Topological interface states can exist in the simple lattice of multi-layer graphene sheets due to the existence of the topological phase transition. Considering an interface, truncated at the middle of two adjacent graphene sheets as schematically shown in figures 3(a) and 3(b). In the left side, the semi-infinite ambient medium has a permittivity of $\varepsilon_a$ (can be dielectric or plasmonic). The plasmonic medium can be doped semiconductors in the infrared and THz regimes [21], described by the Drude-type permittivity $\varepsilon = 1 - \omega_p^2/\omega^2$, where $\omega_p$ is the plasma frequency. Here we set $\hbar\omega_p = 0.3 E_F$ without loss of generality. Note that the interface cutting is chosen at the inversion center of the unit cell. The wave impedance $Z_s$ is defined by the ratio of $E_{//}$ to $H_{//}$ at the interface. For the semi-infinite simple lattice of graphene sheets, the wave impedance at the truncated interface for the TM mode can be derived analytically by the transfer matrix method [6, 17]:

$$Z_s = \frac{k_z}{\omega \varepsilon_0 \varepsilon_h} \cot(qd/2) \tan(k_z d/2) . \quad (6)$$

As a result, $Z_s$ is a function of $k_x$ and $q$. The projected band structures and $Z_s$ are shown in figure 3(c). The

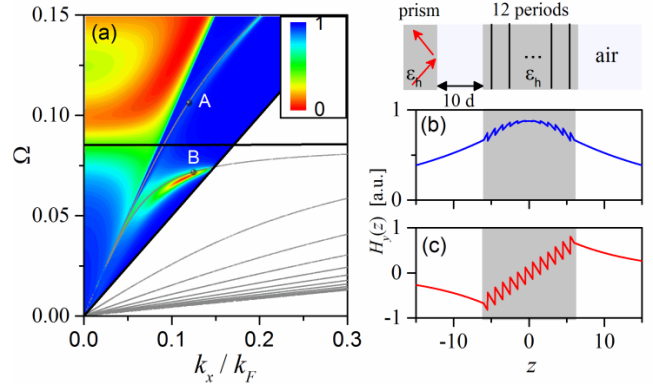

**Figure 4.** Dispersion for the 12-period graphene sheets. The geometry of the structure is sketched in the upper right inset. The ambient dielectric medium is the free space. The gray lines in (a) are the dispersion for all the modes of this system. To excite the interface state, a prism with a permittivity $\varepsilon=\varepsilon_h$ is introduced in the left side. The incident electromagnetic wave is TM polarized. The reflectance as a function of $\Omega$ and $k_x$ of this finite structure is shown in the color-scale form. The loci of the reflection dip agree well with the calculated dispersions. The distribution of the magnetic field $H_y$ for two states marked by A and B are shown in (b) and (c), respectively. The two black solid lines represent the bulk band with $q=0$ for the infinite periodic array of graphene sheets. The parameters $\varepsilon_h$ and $d$ are the same as those in figure 1.

__________________________________________________

blue regions represent the two projected pass bands, and the corresponding wave impedance $Z_s$ are real. The region I, II and III stand for forbidden band gaps, and the corresponding $Z_s$ are purely imaginary.

It can be directly obtained from Eq. (6) that Im$(Z_s)<0$ for region I, whereas Im$(Z_s)>0$ for region II and III. The topological phase transition takes place at the degeneracy point $(k_c, \Omega_c)$ and also gives rise to a discontinuous jump of the sign of Im$(Z_s)$. It is well known the wave impedance for the semi-infinite ambient medium is $Z_a=$Im$(k_z)/\omega\varepsilon_a$ for the states outside the light cone. There should exist an interface state if the condition Im$(Z_s)$+Im$(Z_a)=0$ is satisfied [6]. The dispersion for interface states, solved by this condition, is shown by the red and dark yellow curves in figure 3(c) for dielectric ($\varepsilon_a=1$) and plasmonic medium, respectively. The most striking property is that the degeneracy point $(k_c, \Omega_c)$ acts as a starting or ending point for the dispersion of interface states. For the dielectric ambient medium, the dispersion starts from the near-zero frequency and terminated at the degeneracy point (red curve), while for the plasmonic ambient medium, it starts from the degeneracy point (dark yellow curve), manifesting the non-trivial topological property distinct from the surface plasmon mode on the surface of a Drude metal.

In figure 4, the dispersion for a finite structure with 12-period graphene sheets instead of the semi-infinite structure is shown. The geometry of this structure is sketched in the upper right inset. Here the ambient medium is the free space. There exist two truncated interfaces for this finite system. As expected, there are two branches of dispersion for the interface states lying between the light line $k_x=k_0$ (free space) and $k_x = \sqrt{\varepsilon_h} k_0$ (host medium). Here, the absorption of graphene is taken into account with a finite relaxation time $\tau = \mu E_F / e v_F^2$, where $\mu$ is the DC mobility and $v_F$ is the Fermi velocity [11], e.g., $E_F$=0.6 eV, $\tau$=0.6 ps. The loci of these reflection dips almost coincide with the calculated dispersions. To evaluate the symmetry properties of the two branches of interface states, the distribution of the magnetic field $H_y$ for two states marked by A and B are shown in (b) and (c), respectively. In (b), it can be observed that state A is an even mode, which corresponds to the fundamental waveguide mode of the whole finite system. However, in (c), the topological interface state B is an odd mode and decays exponentially away from the interfaces. For comparison, the bulk band with $q$=0 for the infinite periodic array of graphene sheets are plotted by the two black solid lines. It can be observed that the dispersion corresponding to state A gets into the upper pass band. However, state B locates in the band-gap region I in figure 3(c), starts from the near-zero frequency, and eventually enters the bulk band slightly below the degeneracy point. As we increase the periods of graphene sheets, the dispersion of this interface state converges to that of the interface states for the semi-infinite case (red line in figure 3(c)).

## 4. Conclusions

In summary, topological properties for the multi-layer graphene system have been investigated systematically. There exists a topological phase transition for the TM mode around the accidental degeneracy point, stemming from the hybridization of plasmonic and photonic modes. Band inversion occurs as the wave vector $k_x$ passes through the critical value $k_c$. The Berry phase around this transition point is π, manifesting the non-trivial topological characteristics. Meanwhile, the wave impedance inside the band-gap zones are derived analytically, and a branch of topological interface modes exist as the interface cutting is introduced. The topological phase transition and interface states arise from hybridization of plasmonic and photonic modes can be found in other frequency regime as long as the surface modes exist, which may open up new avenues in both physics and applications for the emerging concept of topological photonics.

## Acknowledgements

This work is supported the National Natural Science Foundation of China (Grant No. 11574037) and the Fundamental Research Funds for the Central Universities (Grant No. CQDXWL-2014-Z005 and No. 106112016CDJCR301205). J. Z. is acknowledge the support from the NSFC. Work in Hong Kong is supported by Hong Kong Research Grant Council AOE/P-02/12. We thank Prof. Z. Q. Zhang and X. B. Jin for helpful discussions.